# 以輔助損失函數引導注意力機制之端對端語者自動分段標記技術

# Speech-Aware Neural Diarization with Encoder-Decoder Attractor Guided by Attention Constraints

李佩穎 Pei-Ying Lee, 郭浩雲 Hau-Yun Guo, 陳柏琳 Berlin Chen
國立臺灣師範大學資訊工程學系
Computer Science and Information Engineering, National Taiwan Normal University
E-mail: {60947089s, 40947006s, berlin }@ntnu.edu.tw

## 摘要

帶有基於編碼器-解碼器吸引子的端到端神經網路（End-to-End Neural Diarization with Encoder-Decoder based Attractor, EEND-EDA）是一種以端對端神經網路為基礎的語者自動分段標記模型，藉由估計吸引子（Attractor）以允許動態辨識不定數量語者。雖然 EEND-EDA 能突破語者數量的限制，卻無法準確捕捉語者的活動資訊，本研究提出了一輔助損失函數（Auxiliary loss），藉由訓練時期的語者活動資訊，引導 EEND-EDA 模型中較低層的 Transformer 編碼器（Transformer Encoder），增強自注意機制（Self-attention）的權重，進而增強對語者動態的關注，我們的實驗結果在公開資料集 Mini LibriSpeech 的評估結果上展現了有效性，將 DER 錯誤率從 30.95%降低至 28.17%。我們將把原始碼公開於 GitHub 網站，供重現研究成果。

**關鍵詞**：語者自動分段標記，端對端神經網路，自注意力機制，語音處理

## Abstract

End-to-End Neural Diarization with Encoder-Decoder based Attractor (EEND-EDA) is an end-to-end neural model for automatic speaker segmentation and labeling. It achieves the capability to handle flexible number of speakers by estimating the number of attractors. EEND-EDA, however, struggles to accurately capture local speaker dynamics. This work proposes an auxiliary loss that aims to guide the Transformer encoders at the lower layer of EEND-EDA model to enhance the effect of self-attention modules using speaker activity information. The results evaluated on public dataset Mini LibriSpeech, demonstrates the effectiveness of the work, reducing Diarization Error Rate from 30.95% to 28.17%. We will release the source code on GitHub to allow further research and reproducibility.

**Keywords**：Speaker Diarization, End-to-End Neural Network, Self-attention mechanism, Speech Processing

## 1. 緒論

是誰在何時說話？[1]想回答此問題，必須利用語者自動分段標記（Speaker Diarization）技術。語者自動分段標記任務旨在分析並標記一段錄音中各語者的活動，擁有豐富的潛在應用，例如：產生帶有語者標籤的會議逐字稿、分析電話會議與會者的發言頻率，此外也有研究 [20]利用語者自動分段標記技術提升語音辨識系統的強健性。

傳統的語者自動分段標記系統仰賴聚類演算法（Clustering Algorithm）運作，例如使用 k-means、AHC（Agglomerative Hierarchical Clustering）等方法將 d-vector、i-vector 等嵌入（Embedding）聚合為語者中心。這些系統在單純的環境下通常擁有良好的表現，然而卻面臨一些難以克服的障礙，例如：以聚類演算法為基礎的系統通常需將輸入音訊切割成多個片段，透過語者活性檢測（Voice Activity Detection, VAD）篩選出存在語音的片段，再從中提取嵌入特徵，這隱含了每個片段至多僅有一位語者

的假設，因此在含有插嘴、搶話等重疊語音的場景往往表現不佳。

為了克服上述的困境，Fujita 等人 [2]提出了 EEND（End-to-End Neural Diarization）架構，以一套端對端神經網路直接預測語者活動，EEND 運用雙向長短期記憶網路（Bi-directional Long-Short Term Memory, Bi-LSTM）對輸入音訊編碼，輔以 PIT（Permutation Invariant Training）解決訓練時期語者排列的問題。EEND 在多項評測基準中創下亮眼的表現，尤其是傳統系統無法解決的重疊語音環境。

目前已經有多個延伸 EEND 架構的相關研究 [3-5]，其中基於編碼器–解碼器的吸引子（Encoder-Decoder Attractor, EDA） [6, 7]，此架構是根據嵌入序列（Embedding sequence）為每個語者生成一個吸引子向量（Attractor vector），然後再進行測量嵌入序列和吸引子向量之間的相似性來生成語者活動，在此模型中使用在二元交叉熵（Binary Cross Entropy, BCE）損失之間的語者標籤和使用無關排列訓練方法從最後一層獲得的預測進行訓練 [5]，達到效能有明顯提升，但是隨著 Transformer 層深度的增加，EEND-EDA 模型的較低區塊與較高區塊相比較於分段標記上的貢獻上相對較小。

為了解決這個問題，[8] 提出在連續的 Transformer 層之間增加 Residual connection，並將輔助 二元交叉熵損失應用於中間層，此模型被修改後的架構稱為 RX-EEND [8]，並且與 SA-EEND 相比，效能有所提高，啟發了我們想從不同角度切入來提高 EEND-EDA 的正確性，根據 [15]中提出的觀察結果，即 Transformer 層的注意力權重矩陣在第3個和第4個矩陣與單位矩陣（Identity Matrix）相似，可以有更多的形式組合，我們假設這對於語者自動分段標記的任務來說是多餘的，並且提出在多語者情況下的輔助損失函數（Auxiliary Losses），對於實現良好的語者自動分段標記表現都很重要 [9-11]，此方法能夠有效地引導自注意力機制，使注意力權重中表現出更多樣的形式組合，強化其捕捉語者活動的能力。

## 1.1. 端對端神經分段標記

在 EEND 模型中將語者自動分段標記問題敘述為每音框多標籤分類任務 [2]，在本研究中，我們將 $X \in \mathbb{R}^{D \times T}$ 表示為 D 維度且長度為 T 的序列（Sequence）音訊特徵，神經網路接受 X 並產生相同長度的語音標籤後驗序列 $Y \in [0,1]^{C \times T}$，其中 $C$ 是語者數量，$[Y]_{c,t}$ 是第 $c$ 位語者在時間 $t$ 時說話的機率，該網路經過訓練後，可以最小化標準答案語者標籤 $Y^* \in \{0,1\}^{C \times T}$ 和預估標籤後驗 Y 之間的二元交叉熵損失：

$$H(y^*, y) = y^* \log y - (1 - y^*) \log(1 - y), \quad (1)$$

$$\mathcal{L}_{Diar}(Y^*, Y) = \frac{1}{CT} \min_{\phi \in \mathcal{P}(C)} \sum_{c=1}^{C} \sum_{t=1}^{T} H([Y^*]_{\phi_c, t}, [Y]_{c,t}), \quad (2)$$

其中 $H(\cdot, \cdot)$ 為二元交叉熵，$\mathcal{P}(C)$ 是序列 $\{1, \cdots, C\}$ 所有可能的語者排列集合，此無關排列的情況 [12, 13]可以處理任意順序語者索引導致的標籤混淆（Label ambiguity）問題。

## 1.2. 帶有基於編碼器-解碼器吸引子的端到端神經網路

EEND 神經網路由一個 Transformer 編碼器（Transformer Encoder）堆疊組成，如 [3,6]所述：

$$E_l = EncoderLayer_l(E_{l-1}) \in \mathbb{R}^{D \times T}, \quad (1 \leq l \leq N), \quad (3)$$

其中 $N$ 是 Transformer 層的數量，初始值為 $E_0 = X^1$，而原始的 EEND [2,3]只需要一個線性層（Linear layer）和一個 S 型函數（Sigmoid function）將 $E_N$ 轉換為 Y，而每個 Transformer 編碼器層由多個自注意力（Multi-head self-attention, MHSA）和完全前饋神經網路（Fully connected feed-forward network）組成，使用縮放點積（Dot-product） [15] 來產生考慮全域特徵關係（Global feature relations）的音框級注意力權重（Frame-level attention weights）：

$$Att(Q, K, V) = Softmax\left(\frac{QK^\top}{\sqrt{d}}\right)V = WV, \quad (4)$$

表 1. Mini LibriSpeech 資料集固定語者的分析

| 資料集 | 音檔數 | 總時長(小時) | 重疊率(%) |
|---|---|---|---|
| train | 500 | 34.45 | 60.49 |
| dev | 500 | 21.08 | 47.86 |

其中 $W \in \mathbb{R}^{T \times T}$ 是考慮全域特徵關係的注意力權重矩陣（Attention weight matrix），定義為 $Q$ 和 $K$ 的乘積除以 $d$ 的平方根，$V \in \mathbb{R}^{T \times d}$ 表示值和 $d$ 分別是隱藏空間（Hidden Space）的維度。

在 EDA [6]中，首先生成特定語者的吸引子向量（Speaker-wise attractor vectors）$A = [a_1, \cdots, a_c] \in \mathbb{R}^{D \times C}$：

$$a_1, \cdots, a_c = EDA(E_N), \quad (5)$$

下列式子表示，EDA 模型使用長短期記憶 (LSTM) 層：

$$(\mathbf{h}_t, \mathbf{c}_t) = LSTM_{enc}(\mathbf{h}_{t-1}, \mathbf{c}_{t-1}, [E_N]_{:,t}), \quad (1 \le t \le T), \quad (6)$$

$$(\mathbf{a}_c, \mathbf{d}_c) = LSTM_{dec}(\mathbf{a}_{c-1}, \mathbf{d}_{c-1}, 0), \quad (1 \le c \le C), \quad (7)$$

其中 $LSTM_{enc}()$ 是單向 LSTM 層，按照順序讀取時間 $t$ 的嵌入向量，$[E_N]_{:,t}$ 是 $E_N$ 的第 $t$ 列的嵌入向量，$\mathbf{h}_t \in \mathbb{R}^D$ 是隱藏狀態（Hidden state），$c_t \in \mathbb{R}^D$ 是單元狀態（Cell state），$LSTM_{dec}()$ 是另一個單向 LSTM 層，初始隱藏狀態 $\mathbf{a}_0 = \mathbf{h}_T$ 和初始元件狀態 $\mathbf{d}_0 = \mathbf{c}_T$，$LSTM_{enc}$ 得到 $C$ 次零向量生成 $C$ 個特定語者的吸引子向量。

EEND-EDA 將嵌入序列 $E_N$ 與特定的語者吸引子向量 A 進行比較來預估語者標籤：

$$Y = Sigmoid(A^T E_N), \quad (8)$$

由於吸引子向量的數量可能會因為公式 7 中的迭代次數 $C$ 而產生變化，EDA 可以共同預估迭代次數來處理未知的語者數量，在本研究中，我們將迭代次數固定為 2，因為我們只在兩個語者對話中評估該方法。

在下一小節中，我們將考慮增加輔助損失函數，來降低語音解析錯誤率。

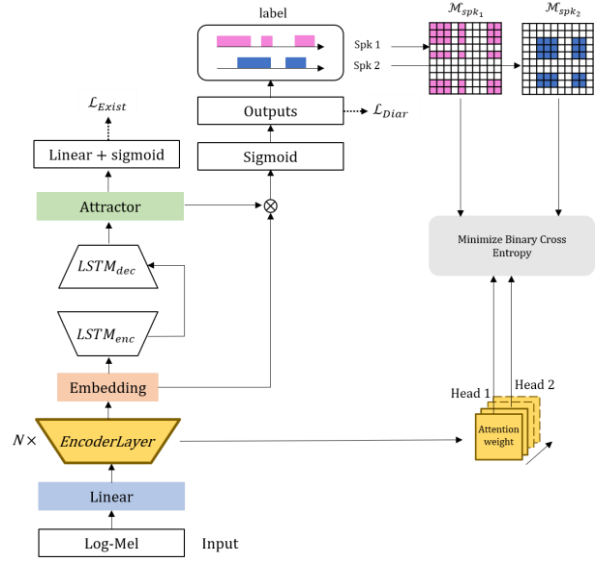

圖 1.加上輔助損失的 EEND-EDA 模型在兩個語者

## 2. 增加輔助損失函數

### 2.1. 語者活動的引導性損失函數

為了引導自注意力機制學習更多不同語者的語音活動形式，本研究加入一輔助損失函數（Auxiliary Loss Function）以導引注意力權重。我們將語者標籤序列 $Y_{\phi_c}$，轉換為目標遮罩（Target masks）$\mathcal{M}_c = Y_{\phi_c}^T Y_{\phi_c} (1 \le c \le C)$：

$$\mathcal{L}_{VAD} = \sum_{c=1}^{C} \mathcal{L}(\mathcal{M}_c, A_c^h) \quad (9)$$

$$= \sum_{c=1}^{C} \frac{1}{T^2} \sum_{i=1}^{T} \sum_{j=1}^{T} H(m_{ij}, a_{ij}), \quad (10)$$

其中 $\mathcal{M}_c$ 所選擇的注意力權重矩陣 $A_c^h$ 是根據第 $h$ 個自注意力計算，而 $m_{ij}$ 和 $a_{ij}$ 分別為 $\mathcal{M}_c$ 和 $A_c^h$ 的第 $(i, j)$ 個項目，因為語者自動分段標記的目標是為每個語者產生 VAD 標籤，因此預期藉由目標遮罩 $\mathcal{M}_c$ 的形式能夠提升自注意力機制的學習成效。

### 2.2. 將輔助損失函數訓練在模型上

因為假設單位矩陣（Identity matrix）是多餘的，所以會根據權重矩陣最相似單位矩陣的自注意力，來應用提出的輔助損失函數，為實現這種方法，將在每個陣列的注意力權重矩陣計算 trace 後降冪排序，選出較無作用的 head。

將分段標記損失 $\mathcal{L}_{Diar}$、輔助損失 $\mathcal{L}_{VAD}$ 以及

表 2. 使用 Mini Librispeech 資料集的評估結果

| System | Loss | α | DER | Miss | FA | Conf. |
|---|---|---|---|---|---|---|
| EEND-EDA | ✗ | N/A | 30.95 | 4.13 | 25.36 | 1.46 |
| EEND-EDA (ep200) | ✗ | N/A | 29.79 | 2.92 | 25.77 | 1.10 |
| EEND-EDA + VAD Loss (ep50) | ✓ | 0.08 | 30.70 | 3.03 | 26.55 | 1.11 |
| EEND-EDA + VAD Loss（本研究） | ✓ | 0.08 | **28.17** | **2.91** | **24.35** | **0.91** |
| EEND-EDA + VAD Loss* | ✓ | 0.08 | 30.08 | 1.75 | 23.61 | 1.75 |

吸引子損失 $\mathcal{L}_{Exist}$，一起幫助每個陣列不僅可以區分語音的存在也可以區分每個語者的語音，最後，我們把總損失函數定義為：

$$\mathcal{L} = \mathcal{L}_{Diar} + \alpha\mathcal{L}_{VAD} + \beta\mathcal{L}_{Exist} \quad (11)$$

其中 α 和 β 為超參數。

### 2.3. 相關研究

本研究與 [14]探討了類似的議題，該研究採用 SA-EEND 模型 [3]，針對 Transformer 編碼器層進行了改進，提出在 Transformer 編碼器層的自注意力機制中使用輔助損失，成功增加模型訓練上的效果，我們的研究不同之處在於應用在 EEND-EDA 模型上，有助於未來應用在多語者的會話、動態語者數量的場景進行研究，解決模型的訓練表現。

## 3. 資料集與實驗設定

### 3.1. 資料集

本研究使用的資料集 Mini LibriSpeech 是一套公開的模擬資料集，其內容是從 LibriSpeech 選取部分錄音，與 MUSAN [19]資料集中的噪音混和疊加以模擬兩位語者對話的情境。我們使用與 ESPNET 完全相同的程式[1]來建置資料集，以提供可復現、易於比較的實驗數據。

表 1 為該資料集的統計分析資料，在訓練（train）資料集和驗證（dev）資料集中各有 500 個音檔，每個音檔都有不特定的兩位語者進行對話，在訓練資料集中，總說話時長為 34.45 小時，在兩位語者同時對話的重疊率為 60.49%，在驗證資料集中，總說話時長為 21.08 小時，且在兩位語者同時對話的重疊率為 47.86%。

### 3.2. 實驗設定

EEND-EDA：我們使用了 [16]中以 PyTorch 實作的 EEND-EDA 模型作為基準系統，除特別說明以外，模型設定與 [16]完全相同，也與 [3, 6]一致。輸入的聲音訊號係經以下步驟轉換為對數梅爾頻譜：聲音訊號的取樣率設定為 8000 Hz，接著以 25 ms 為單位，彼此重疊 10 ms 的序列通過 23 維的梅爾濾波器組（Mel Filterbank），再對結果取對數。每個音框與前後各七個組成特徵後，以每十組特徵為間隔再取樣一次。因此，模型的輸入特徵為 345 (23×15) 維的序列，間隔為 100 ms。值得一提的是，本研究中以 TorchAudio [17]取代了 librosa 的聲音訊號處理模組，但並未改變音訊特徵的處理流程。

此模型利用自注意機制（Self-attention）來擷取音框的嵌入資訊（Embedding），每層 Transformer 編碼器具有 256 個隱藏單元及四個 head，共有四層，其中 $N$ 設定為 4。

此模型的訓練批次樣本數設為 48，共訓練了 100 epochs，使用 Adam 最佳化器（Optimizer）調整參數，並以 [15]中提到的 Noam 排程法調整學習率，學習率起步（Warm up）的步數設定為 10000 步，訓練過程中最高達到的最高學習率為 $3.62 \times 10^{-4}$。

EEND-EDA + VAD Loss：以基準系統為基礎，再其訓練完成後，加入了公式 10 描述的輔助損失函數，額外訓練了 100 epochs，α 和 β 值分別設為 0.008 及 1.0。計算輔助損失函數時，係以第四層

---
[1] https://github.com/espnet/espnet/blob/master/egs2/mini_librispeech/diar1/local/data.sh
*加入 VAD Loss 訓練時，將學習率固定於 1e-5

Transformer 編碼器為對象。VAD Loss 僅作用於模型訓練時期，不會增加推論時期的複雜度。

### 3.3. 效能評估

本研究使用 [18] 提出的語音解析錯誤率（Diarization Error Rate, DER）來評估系統效能，其定義為：

$$\text{DER} = \frac{T_{MI} + T_{FA} + T_{CF}}{T_{Speech}} \quad (12)$$

其中 $T_{MI}$、$T_{FA}$、$T_{CF}$ 和 $T_{Speech}$ 分別表示為，漏報（Miss）、誤報（False alarm, FA）、混淆（Confusion, Conf.）和錄音總時長，並且在每 $e$ 個語音邊界處使用 0.25 的容許值（Collar）進行評估。

### 3.4. 實驗結果

表 2 列出了基準系統與本研究改進的系統效能比較。為進一步探究訓練 Epoch 值對模型表現的影響，我們也對基準系統額外訓練 100 Epochs（EEND-EDA (ep200)）以進行公平的對比，EEND-EDA + VAD Loss (ep50) 則提供了加入輔助損失函數後，模型尚未收斂前的測試數據。雖然經過更多時間的訓練後，能幫助 EEND-EDA 模型收斂更完全，但改善幅度微小，仍不如加入 VAD 輔助損失函數來的有效，結果顯示本研究提出之方法的確能應證假說，提供優於 EEND-EDA 系統的表現，將語音解析錯誤率從 30.95%降低至 28.17%。

## 4. 結論

本研究提出了基於語者標籤的輔助損失函數，以引導自注意機制更關注具有語者活動的片段，在 EEND-EDA 模型上的研究結果顯示加入輔助損失函數訓練可以有效改善模型表現，呼應了 [14]的觀察。此外我們分析了以不同超參數的設定，使模型針對 Mini LibriSpeech 資料集收斂地更好。本研究原始碼將公開於 GitHub 網站。未來我們將針對不同資料集和更多語者的會話、動態語者數量的場景進行研究。